\documentclass[aip,pof,amsmath,amssymb,reprint,superscriptaddress]{revtex4-2}

\usepackage{graphicx}
\usepackage{bm}
\usepackage{booktabs}
\usepackage{tikz}
\usetikzlibrary{arrows.meta,positioning,calc,fit,backgrounds}
\usepackage[dvipsnames]{xcolor}

\definecolor{cnonact}{HTML}{009988}
\definecolor{copp}{HTML}{0077BB}
\definecolor{cgru}{HTML}{EE7733}
\definecolor{ces}{HTML}{CC3311}

\newcommand{\Retau}{\mathit{Re}_{\tau}}
\newcommand{\ufric}{u_{\tau}}
\newcommand{\Ubulk}{U_{b}}
\newcommand{\ww}{w_{w}}
\newcommand{\tauw}{\tau_{w}}
\newcommand{\zd}{z_{d}}
\newcommand{\zp}{z^{+}}
\newcommand{\DR}{\mathrm{DR}}
\newcommand{\Epump}{\mathcal{P}_{p}}
\newcommand{\Ww}{\mathcal{W}_{w}}
\newcommand{\Eps}{\varepsilon}
\newcommand{\avg}[1]{\left\langle #1 \right\rangle}

\newcommand{\figmaybe}[2][\linewidth]{%
  \IfFileExists{#2}{\includegraphics[width=#1,height=0.78\textheight,keepaspectratio]{#2}}%
  {\fbox{\parbox[c][2.6cm][c]{#1}{\centering\itshape figure pending: \detokenize{#2}}}}}
\newcommand{\fignat}[1]{%
  \IfFileExists{#1}{\includegraphics[scale=1.0]{#1}}%
  {\fbox{\parbox[c][2.6cm][c]{0.6\linewidth}{\centering\itshape figure pending: \detokenize{#1}}}}}

\begin{document}

\title{Gradient-free learning of a closed-loop wall controller for
       turbulent drag reduction}

\author{G.~M. Cavallazzi}
\email{giorgio.cavallazzi@city.ac.uk}
\author{M. P\'erez Cuadrado}
\author{A. Pinelli}
\affiliation{Department of Engineering, City St.\ George's, University of London,
             Northampton Square, EC1V 0HB, London, United Kingdom}

\begin{abstract}
Closed-loop wall controllers learnt by multi-agent reinforcement learning are
usually trained on periodic boxes far smaller than the flows they are meant to
drive, and a large part of their drag reduction is lost when they are carried
across. Retraining on the target domain is not an affordable remedy: the centralised critic that assigns credit to each wall patch
degrades as patches are added, the zero-net-mass constraint couples the patches it
is asked to separate, and the episodes must be collected in sequence at a cost
that grows with the domain. We propose instead a short gradient-free refinement
stage, applied to the transferred policy on the domain it will drive. An evolution
strategy scores whole flow episodes against a regularised objective, so no credit
has to be assigned to individual patches, and the candidates in a generation
are independent and run in parallel. Applied to a recurrent multi-agent
policy trained on a minimal flow unit at $\Retau\simeq180$ and evaluated on a
channel sixteen times larger in wall-parallel area, a few generations raise the
drag reduction from $19.0\%$ to $25.7\%$, above the $22.5\%$ of opposition
control. Because the policies before and after
refinement share an architecture and a training history, the difference between
the controlled flows follows from the refinement alone. It shows in the friction
decomposition, in the Reynolds stresses and in the near-wall spectra, and the
actuation moves from a weak
coupling to the wall-normal velocity towards a strong coupling to the streamwise
fluctuation.
\end{abstract}

\maketitle

\section{Introduction}\label{sec:intro}

Skin-friction drag in wall-bounded turbulence is generated by quasi-streamwise
vortices whose induced wall-normal velocities sustain the Reynolds shear stress
above its laminar value.\cite{orlandi1994generation} The vortices are part of a
self-sustaining near-wall cycle of low-speed streaks and quasi-streamwise
vortices\cite{kline1967structure,jimenez1991mfu,hamilton1995regeneration,waleffe1997ssp}
that persists autonomously within the buffer layer even in the absence of
outer-flow forcing,\cite{jimenez1999autonomous} which makes that layer the natural
target for drag-reduction control.

Established active methods act on the cycle through a specific and physically
motivated interaction. Opposition control opposes the wall-normal velocity sensed
at a plane inside the buffer layer,\cite{choi1994opposition,hammond1998oppositionmech}
suboptimal control derives a feedback law from the suppression of the near-wall
Reynolds shear stress,\cite{lee1998suboptimal,fukagata2004suboptimal} resolvent
analysis proposes control laws from the dominant response modes of the linearised
operator,\cite{kawagoe2019resolvent} and spanwise forcing and streamwise-travelling
waves reorganise the streaks laterally.\cite{quadrio2009stw,gatti2016stwhighre,touber2012spanwise}
In each case the actuation is designed around the interaction from the outset.
Controllers discovered by data-driven optimisation carry no such prior, and what
physical interaction they exploit is not known in advance.

Reinforcement learning has been the most productive route to such controllers.
Applied to the plane channel at low Reynolds number it has produced closed-loop
wall policies reporting drag reductions competitive with the physics-based
benchmarks,\cite{guastoni2023drl,sonoda2023rl,lee2023drlchannel,han2020cnnchannel,zhou2025drlhighre,varela2022actuators,cavallazzi2024wallcycle}
as surveyed in several reviews.\cite{garnier2021reviewdrl,brunton2020mlreview,vignon2023review}
A common way to keep the formulation tractable is a shared, translationally
invariant multi-agent policy\cite{vignon2023rb,guastoni2023drl} trained with a
centralised critic on a small periodic domain, executed independently at each wall
patch, and then applied to a larger channel. This template underlies numerous
subsequent studies.\cite{sonoda2023rl,cavallazzi2024wallcycle,walchli2024minimalchannel,bae2022wallmodel,suarez2024marl3dcylinder,rabault2019multienv}
Evolutionary optimisation reached the same problem earlier and by a different
route, through genetic algorithms and genetic programming acting on
low-dimensional sets of actuation
parameters,\cite{gautier2015separation,zhou2020aijet,yu2026gatbl,kasagi2009mems}
and through covariance-matrix adaptation applied to the material parameters of a
compliant surface.\cite{fukagata2008compliant}

That last study is worth dwelling on, because it contains in miniature the
difficulty the present work addresses. Fukagata \textit{et al.}\ optimised an
anisotropic compliant wall by an evolution strategy in a channel of
$3\delta\times2\delta\times3\delta$ and obtained about $8\%$ drag reduction; when
the optimised wall was then placed in a domain twice as long in the streamwise
direction, the drag increased instead.\cite{fukagata2008compliant} The optimiser
had found a wall that suited the box rather than the flow, and the box had
suppressed the large-scale two-dimensional structures the wall would go on to
amplify.\cite{luhar2016compliant} The same hazard attends any controller
optimised in a domain smaller than the one it will drive, and it attends learnt
policies with particular force, because a policy evaluated in a minimal flow unit
can push the flow towards relaminarisation, at which point the pressure gradient
falls to its laminar value and the reported drag reduction saturates at a number
that reflects the box rather than the
controller.\cite{cavallazzi2026rewardhacking} A representative figure therefore
requires a domain large enough to keep the flow fully turbulent throughout, which
is the setting adopted here.

Training, however, is confined to a small domain by the structure of the
multi-agent update and not by simulation cost alone. A gradient step adjusts the
shared policy parameters from a per-patch return whose credit is assigned by a
central critic, and that critic must resolve each patch within a joint action
space that widens as patches are added, so the estimate deteriorates on larger
domains. Two features of the wall-control problem tie the restriction down, one
corrupting the credit assignment and the other raising its cost. Because the
actuation must inject no net mass, the joint action is projected onto its
zero-mean subspace before reaching the wall, and under that projection the action
applied at any single patch depends on all the others, which breaks the
separability of per-patch credit.\cite{cavallazzi2026rewardhacking} Every training
episode also requires advancing a costly direct numerical simulation, so the
wall-clock cost of collecting experience grows with the domain. Training is
consequently restricted to a minimal flow unit and the policy is transferred only
afterwards, a pattern that has a long history in machine learning generally, where
features learnt on one distribution are reused on
another,\cite{yosinski2014transferable,taylor2009transferrl} and where the gap
between the training and the deployment distribution is closed either by
randomising the source\cite{tobin2017domainrand,peng2018simtoreal} or by
continuing to train on data the deployed system itself
generates.\cite{ross2011dagger,um2020solverloop}

An evolution strategy closes that gap directly. It treats the expected return as a
function of the policy parameters, evaluates a population of perturbed parameter
vectors over complete flow episodes, and updates the mean from a weighted average
of the perturbations, with no backpropagation through the flow and no central
critic.\cite{rechenberg1973evolution,hansen2001cmaes,wierstra2014nes,salimans2017es}
Because whole episodes are scored, the per-agent credit assignment that the
zero-net-mass constraint corrupts never arises. The population evaluations are
independent and parallelise across rollouts. The search can therefore run on the
large evaluation domain itself, with no small-box surrogate, and the natural way
to use it is not as a replacement for gradient-based training but as a second
stage after it: the minimal box is cheap and locates a good region of parameter
space, and the strategy refines within that region on the domain that matters.

This paper carries out that two-stage procedure and asks what it buys. We take a
recurrent multi-agent policy trained on a minimal flow unit at
$\Retau\simeq180$,\cite{cavallazzi2026rewardhacking} refine its parameters by an
evolution strategy on a channel sixteen times larger in wall-parallel area, and
compare the flows before and after refinement against opposition control and the
non-actuated channel. Because the two learnt controllers share an architecture and
a training history and differ only by the refinement stage, the comparison
isolates what refinement on the target domain changes, both in the flow and in the
control law. We characterise the controlled states with a wall-normal
decomposition of the skin friction,\cite{fukagata2002fik} Reynolds-stress profiles
scaled by each controller's own friction velocity, the anisotropy invariants of
the stress tensor,\cite{lumley1977isotropy} the premultiplied near-wall
spectra,\cite{delalamo2003vlsm,delalamo2004scaling} and the conditional dependence
of the actuation on the sensed fluctuations.

\section{Problem set-up and methodology}\label{sec:setup}

\subsection{Flow configuration and numerics}\label{sec:numerics}

We perform direct numerical simulations of an incompressible flow in a plane
channel at constant flow rate using CaNS.\cite{costa2018cans,costa2021gpucans} The
solver integrates the unsteady incompressible Navier--Stokes equations with a
low-storage third-order Runge--Kutta advancement within a pressure-projection
algorithm.

Throughout, $x$ is the streamwise direction, $y$ the spanwise direction and $z$
the wall-normal direction, and $u$, $v$ and $w$ are the corresponding velocity
components with $p$ the pressure. The reader is asked to note that $w$ denotes the
wall-normal component here, which is the opposite of the convention used in part
of the literature. The flow is periodic in $x$ and $y$, bounded by a wall at
$z=0$ and closed by a symmetry plane at the channel centreline, so that only the
lower half of the channel is simulated. Primes denote fluctuations about the plane-averaged mean and $\avg{\cdot}$
denotes an average over wall-parallel planes and time.

The friction Reynolds number is $\Retau=H^{+}\simeq180$, with $H$ the channel
half-height. The evaluation domain measures
$(L_x^{+},L_y^{+},L_z^{+})\simeq(1922,576,180)$ in viscous units, equivalently
$(L_x/H,L_y/H,L_z/H)=(10.68,3.2,1)$, discretised on a $256\times256\times100$ grid
with wall-normal points clustered towards the wall, giving
$\Delta x^{+}=7.5$ and $\Delta y^{+}=2.3$. The minimal flow unit on which the
learnt policy was originally trained\cite{cavallazzi2026rewardhacking} shares the
Reynolds number, the wall-normal grid, the wall-parallel resolution, the patch
size in wall units and the actuation cadence, and differs only in extent,
$(L_x^{+},L_y^{+})\simeq(481,144)$ on a $64\times64$ wall-parallel grid. The
transfer studied here is therefore a transfer in domain size alone.

Control is applied as wall-normal blowing and suction $\ww(x,y,t)$ at the wall
$\Gamma_w$, with zero net mass flux enforced at every instant,
$\int_{\Gamma_w}\ww\,\mathrm{d}x\,\mathrm{d}y=0$. The streamwise and
wall-normal velocity fluctuations are sensed at a detection plane at
$\zd^{+}\simeq14$, and they are the only flow quantities the controllers see. The actuation interval is
$\Delta t^{+}\simeq5$, longer than the fastest near-wall fluctuations but short
compared with the typical streak lifetime.

\subsection{Energy budget and performance metrics}\label{sec:budget}

The flow rate is held constant by a uniform body force whose magnitude equals the
instantaneous mean streamwise pressure gradient, so the pressure field solved for
carries no mean-gradient contribution and its volume mean is set to zero. The bulk
kinetic-energy budget then closes on two external power terms, the pumping power
supplied by the flow-rate controller, per unit volume, and the work the actuation
does on the fluid, per unit wall area,
\begin{equation}
  \Epump = -\avg{\partial_x p}\,\Ubulk ,
  \qquad
  \Ww = -\frac{1}{L_xL_y}\int_{\Gamma_w}\avg{\ww\,p}\,\mathrm{d}x\,\mathrm{d}y ,
  \label{eq:powers}
\end{equation}
with $\Ubulk=1$ in code units. The wall power is independent of the pressure
datum, since the zero-net-mass condition makes the wall-integrated actuation
vanish at every instant. The viscous contribution vanishes identically: continuity
at a wall that is no-slip in the wall-parallel components forces
$\partial_z w|_{w}=0$ pointwise. The total expenditure is
$\Eps=\Epump+\Ww/H$, and performance is measured as the net saving
$(\Eps_0-\Eps)/\Eps_0$ relative to the non-actuated channel. The boundary work
also carries a flux of kinetic energy $\tfrac12\ww^{3}$ through the
wall.\cite{choi1994opposition,fukagata2009netpower} It is retained in the budget
reported below and it is small: for the controller of \S\ref{sec:es} it measures
$2.7\times10^{-7}$, four per cent of $\Ww$ and under one part in $10^{4}$ of
$\Epump$.
\begin{equation}
  \DR = 1 - \frac{|\avg{\partial_x p}|}{|\avg{\partial_x p}_0|}
  \label{eq:dr}
\end{equation}
is the bare drag reduction, which tracks the mean pressure gradient and says
nothing about $\Ww$.

Part of the flow-control literature charges the actuation through its amplitude
alone, taking $\sim|\ww|^{3}$ as a surrogate for the wall power. The surrogate is
a function of the actuation amplitude and of nothing else, so it can cap the
output but it cannot register the covariance $\avg{\ww\,p}$ through which a
bounded-amplitude actuation still delivers real power to the flow. The two part
company where it matters. A controller whose actuation saturates can drive the
wall power high enough to raise the total expenditure above the non-actuated value
while the bare drag reduction, and the amplitude surrogate along with it, still
report a saving.\cite{cavallazzi2026rewardhacking,marusic2021energypath} That
regime is excluded here by construction, and its absence is visible in the
numbers: the controller of \S\ref{sec:es} delivers $\Ww=6.9\times10^{-6}$ against
$\Epump=3.0\times10^{-3}$, so the wall power is $0.23$ per cent of the pumping
power and the net saving and the drag reduction agree to within two tenths of a
percentage point.
Agreement between the two is the signature of an actuation that is doing
negligible work on the flow, and it is a result of the evaluation rather than an
assumption of it.

That budget is how the controllers are assessed. It is not what either of them was
optimised on, and the distinction is deliberate. In a gradient-based setting the
reward defines the landscape the search moves through, and substituting the full
budget for the drag-reduction signal changes its shape rather than
sharpening it: the covariance is a single global scalar that responds weakly to
any one patch, while the amplitude terms that would accompany it press on every
patch at once. The objective of \S\ref{sec:es} therefore keeps drag reduction as
the signal and adds regularisation whose role is to hold the search in the region
where the actuation stays smooth, unsaturated and zero in the mean. That is the
region in which the budget and the drag reduction coincide, so the reward guides
the search towards a physically sound equilibrium rather than standing in for the
quantity the evaluation measures.

\subsection{The closed-loop controller}\label{sec:controller}

The controller is a recurrent neural network that maps the velocity field at the
detection plane to a wall-normal velocity field at the wall. The wall is
partitioned into square patches of $4\times4$ grid cells, which is
$30^{+}\times9^{+}$ in wall units, giving $64\times64=4096$ patches on the
evaluation grid and $16\times16=256$ on the minimal flow unit. One shared network
is applied at every patch, so the policy is translation-equivariant and the same
parameters define a controller on any grid.

Each patch observes a $3\times3$ neighbourhood of patches, a $12\times12$ window
of grid cells with periodic wrap, in three channels: the streamwise and
wall-normal velocity fluctuations at the detection plane, normalised by
$\omega_{\max}=0.064$, and the previous action tiled over the window. The window
passes through two convolutional blocks with $16$ and $32$ channels, each followed
by group normalisation, a rectified linear unit and $2\times2$ max pooling, then a
linear projection to $64$ features, a single gated recurrent unit of hidden size
$64$, layer normalisation, and a linear map to one scalar passed through a
hyperbolic tangent. Writing $o_t$ for the observation, $h_t$ for the hidden state
and $a_t$ for the field of per-patch scalars,
\begin{equation}
  (a_t,h_t) = f_{\theta}(o_t,h_{t-1}),
  \qquad
  \ww(\cdot,t) = s\,\mathcal{P}\bigl[G_{\sigma_s} * a_t\bigr],
  \label{eq:policy}
\end{equation}
where $G_{\sigma_s}*$ is a periodic Gaussian smoothing with
$\sigma_s=1.5$ grid cells, $s=0.8\,\omega_{\max}=0.0512$ sets the amplitude, and
$\mathcal{P}:a\mapsto a-\bar{a}$ subtracts the spatial mean. The projection is the
final layer of the network rather than a correction applied afterwards, so
$\int_{\Gamma_w}\ww\,\mathrm{d}x\,\mathrm{d}y=0$ holds at every instant for any
parameter vector.

The parameter vector $\theta\in\mathbb{R}^{d}$ collects every weight and bias of
that network, the two convolutional kernels and their normalisation parameters,
the feature projection, the recurrent unit's input and hidden weight matrices and
biases, the layer normalisation and the output layer. Its dimension is
$d=48\,833$. It is the only object the optimisation of \S\ref{sec:es} acts on.

\subsection{Refinement by an evolution strategy}\label{sec:es}

The parameters are first obtained by multi-agent gradient descent on the minimal
flow unit within a centralised-training, decentralised-execution
framework,\cite{lowe2017maddpg,foerster2018coma} as described
elsewhere.\cite{cavallazzi2026rewardhacking} They are then refined on the
evaluation domain by an evolution
strategy,\cite{rechenberg1973evolution,wierstra2014nes,salimans2017es} which
requires no derivative of the objective with respect to $\theta$. Obtaining one
would mean differentiating through both the flow solver and the recurrent network;
instead each candidate is scored by running one complete flow episode and reading
off the result, with the simulation and the network treated as a black box.

\begin{figure}[tb]
\centering
\resizebox{\linewidth}{!}{%
\begin{tikzpicture}[
  every node/.style={font=\footnotesize},
  >=Stealth, line cap=round,
  box/.style={draw,thick,rounded corners,minimum height=9mm,minimum width=24mm,
              align=center,inner sep=2pt},
  dns/.style={draw,thick,draw=cnonact,fill=cnonact!12,rounded corners,
              minimum height=7.5mm,minimum width=34mm,align=center,inner sep=2pt},
  flow/.style={->,thick,gray!65},
  lbl/.style={font=\scriptsize\itshape}
]
\node[box,draw=ces,fill=ces!12] (mean) at (0,0) {mean $m$\\ scale $\sigma$};
\node[box,draw=copp,fill=copp!12] (sample) at (3.9,0)
  {mirrored sampling\\ $m\pm\sigma\bm{\xi}_k$};
\node[dns] (d1) at (8.6, 1.55) {episode $1$ $\rightarrow$ $F_1$};
\node[dns] (d2) at (8.6, 0.30) {episode $2$ $\rightarrow$ $F_2$};
\node[font=\footnotesize] (dots) at (8.6,-0.72) {$\vdots$};
\node[dns] (dn) at (8.6,-1.75) {episode $2n$ $\rightarrow$ $F_{2n}$};
\node[box,draw=cgru,fill=cgru!15] (rank) at (13.3,0)
  {centred ranks\\ $\upsilon_k$};
\node[box,draw=ces,fill=ces!12] (adam) at (13.3,-3.15)
  {Adam step\\ Eq.~\eqref{eq:es_update}};

\draw[flow] (mean) -- (sample);
\draw[flow] (sample.east) -- (d1.west);
\draw[flow] (sample.east) -- (d2.west);
\draw[flow] (sample.east) -- (dn.west);
\draw[flow] (d1.east) -- (rank.west);
\draw[flow] (d2.east) -- (rank.west);
\draw[flow] (dn.east) -- (rank.west);
\draw[flow] (rank) -- (adam);
\draw[flow] (adam.south) -- ++(0,-0.75) -| (mean.south)
  node[pos=0.28,above,font=\scriptsize,black] {$m \leftarrow m + \text{Adam}(\hat{g})$};

\begin{scope}[on background layer]
\node[draw=cnonact!70,dashed,thick,rounded corners,inner sep=3.5mm,
      fit=(d1)(d2)(dots)(dn),
      label={[lbl,text=cnonact!80!black]above:{independent, evaluated in parallel}}] {};
\end{scope}
\end{tikzpicture}}
\caption{One generation of the evolution strategy. The current mean is perturbed
into $2n=24$ mirrored candidates, each scored by one complete flow episode on the
evaluation domain. The episodes are mutually independent, so a generation costs
one episode of wall-clock time given enough parallel resources. Only the ordering
of the $2n$ fitness values enters the update, through the centred ranks.}
\label{fig:esloop}
\end{figure}
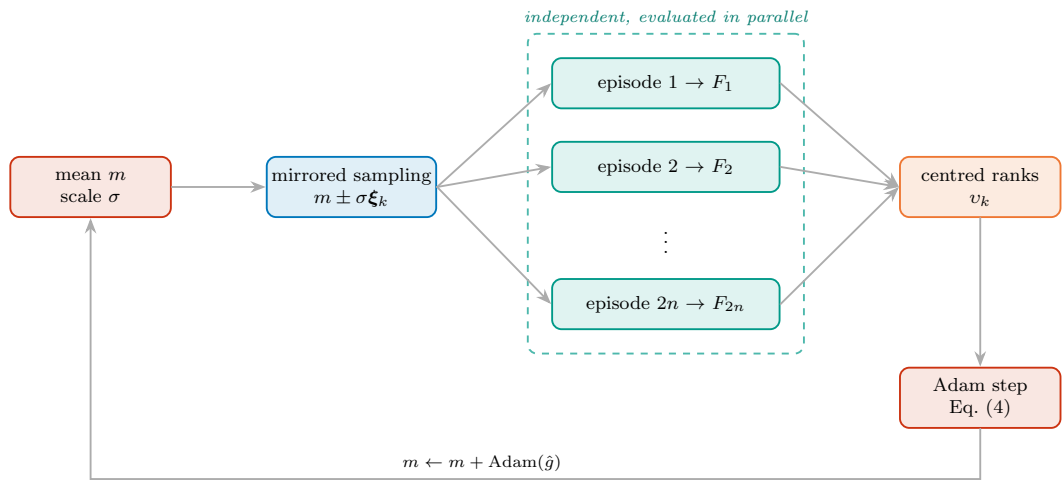

At each generation, $n=12$ mirrored pairs of parameter vectors are drawn by adding
Gaussian perturbations to the current mean $m\in\mathbb{R}^{d}$,
$\{m\pm\sigma\bm{\xi}_k\}_{k=1}^{n}$ with
$\bm{\xi}_k\sim\mathcal{N}(0,\mathbf{I})$ and $\sigma=0.02$, giving $2n=24$
members. Each is evaluated as one episode initialised from the same flow state, so
that fitness differences reflect $\theta$ rather than the initial
condition.\cite{salimans2017es} The mean is advanced along
\begin{equation}
  \hat{g} = \frac{1}{n\sigma}\sum_{k=1}^{n}
            \bigl(\upsilon_k^{+}-\upsilon_k^{-}\bigr)\bm{\xi}_k ,
  \label{eq:es_update}
\end{equation}
where $\upsilon_k^{\pm}$ is the centred rank of the fitness of
$m\pm\sigma\bm{\xi}_k$ within the generation, that is, its position among the $2n$
values shifted so the ranks average to zero. The update therefore depends only on
the relative ordering of the population and not on the numerical scale of the
fitness.\cite{salimans2017es} The mean is updated from $\hat{g}$ using
Adam\cite{kingma2014adam} with a step size of $0.01$ and a weight decay of
$0.005$. Figure~\ref{fig:esloop} sketches the generation.

The fitness is the drag reduction of Eq.~\eqref{eq:dr} less four penalties that
act on the raw per-patch scalars $a$ before smoothing and scaling,
\begin{equation}
  F(\theta) = \DR
    - \lambda_{t}\avg{(\Delta_t a)^{2}}
    - \lambda_{s}\avg{(\Delta_x a)^{2}+(\Delta_y a)^{2}}
    - \lambda_{0}\avg{\bar{a}^{2}}
    - \lambda_{e}\avg{a^{2}} ,
  \label{eq:fitness}
\end{equation}
with $(\lambda_t,\lambda_s,\lambda_0,\lambda_e)=(0.025,0.05,2.0,0.1)$, the same
weights used as auxiliary losses during the gradient-based
stage.\cite{cavallazzi2026rewardhacking} The temporal and spatial terms penalise
actuation that changes abruptly in time or between neighbouring patches, the third
penalises any residual mean, and the fourth bounds the amplitude. They are not
cosmetic. Optimising the unpenalised net energy saving on the same domain drives
the policy onto a large-amplitude travelling wall wave whose wavelength is set by
the streamwise extent of the box rather than by the near-wall cycle, a degenerate
solution that scores well and controls nothing; that outcome is documented in
Appendix~\ref{app:unreg} and is the reason Eq.~\eqref{eq:fitness} takes the form
it does.

Each candidate is scored over a window of $70$ actuation steps, $29.75$ convective
time units, after $8$ steps of settling. One eddy turnover is
$h/\ufric\simeq15.6$ time units, so a candidate is scored over less than two of
them. This is a
short window, and short windows are capable of ranking noise rather than policies;
we return to this in \S\ref{sec:res-transfer}, where the selected candidate is
re-evaluated over $500$ time units and its ranking is shown to hold. The
optimisation ran for eight generations at $24$ candidates each, $192$ episodes and
about $6.4\times10^{3}$ time units of direct numerical simulation in total, or
$19$ hours on a single GPU. The $2n$ episodes within a generation are mutually
independent, so with sufficient parallel resources the wall-clock cost of a
generation reduces to that of a single episode.

Three details of the run are recorded here for completeness. The optimisation was
carried out on a wall-normal grid whose stretching places the sensing index at
$\zp\simeq9.8$, whereas the evaluation grid used throughout
\S\ref{sec:results} places it at $\zp\simeq13.8$, so the refined policy is assessed
at a slightly higher sensing station than the one it was refined at. The deployed
parameter vector is not the one of highest fitness in the run but the one of
substantially lowest penalty at essentially equal fitness, $0.015$ against $0.033$
for a fitness difference of $0.0005$, a choice made to avoid rewarding candidates
that buy drag reduction with rough actuation. The optimiser moments were reset
once, at a scheduler-imposed pause between the second and third generations.

\subsection{Control cases and diagnostics}\label{sec:cases}

Four flows are compared. The non-actuated channel provides the reference.
Opposition control sets the wall velocity to the negative of the wall-normal
velocity fluctuation sensed at the detection plane,
$\ww=-w'(\zd)$.\cite{choi1994opposition} GRU-MARL is the policy of
\S\ref{sec:controller} trained by gradient-based multi-agent learning on the
minimal flow unit and applied to the evaluation domain without further
training.\cite{cavallazzi2026rewardhacking} ES is the same policy after refinement
on the evaluation domain by the strategy of \S\ref{sec:es}. The two learnt
controllers therefore share an architecture, a training history and a sensing
station, and differ only by the refinement stage, which is what allows the
comparison below to attribute differences to that stage. We omit the saturated,
memoryless policies whose behaviour motivates a separate study of reward
specification,\cite{cavallazzi2026rewardhacking} since the concern here is the
mechanism of controllers that reduce drag at low actuation power.

Each controlled flow is characterised by its own friction velocity
$\ufric=\sqrt{\tauw/\rho}$, with $\tauw$ the wall shear stress and $\rho$ the
density, evaluated after the control has acted. All wall-scaled quantities below
are normalised by this flow-specific value, so that a difference between two
scaled profiles is a difference in the shape of the turbulence and not an artefact
of the drag reduction itself.

The contribution of the Reynolds shear stress to the skin friction is quantified
with the identity of Fukagata \textit{et al.},\cite{fukagata2002fik} which
decomposes the friction coefficient $C_f=\tauw/(\tfrac12\rho \Ubulk^{2})$ into a
laminar contribution at the same bulk Reynolds number and a wall-normal weighted
integral of the Reynolds shear stress,
\begin{equation}
  C_f = C_f^{\mathrm{lam}}
        + c\int_{0}^{1}(1-z/H)\bigl(-\avg{u'w'}\bigr)\,\mathrm{d}(z/H),
  \label{eq:fik}
\end{equation}
integrated from the wall to the centreline with the plane-channel prefactor
$c=12$. Scale information comes from the premultiplied spectra of the velocity
fluctuations at fixed $z/H$.\cite{delalamo2004scaling}

\section{Results}\label{sec:results}

\subsection{Transfer loss and its recovery}\label{sec:res-transfer}

On the minimal flow unit the gradient-trained policy sustains a drag reduction of
between $24\%$ and $27\%$, the value recorded for the deployed checkpoint being
$26.8\%$ and the mean over the final fifty training episodes $24.3\%$.
Carried unchanged to the evaluation domain, the same parameters deliver
$19.0\%$.
The gap is not a property of the architecture, which is unchanged, nor of the
resolution, the Reynolds number, the patch size in wall units or the actuation
cadence, all of which are matched between the two domains. It is what the training
box was worth.

\begin{figure}[tb]\centering
  \fignat{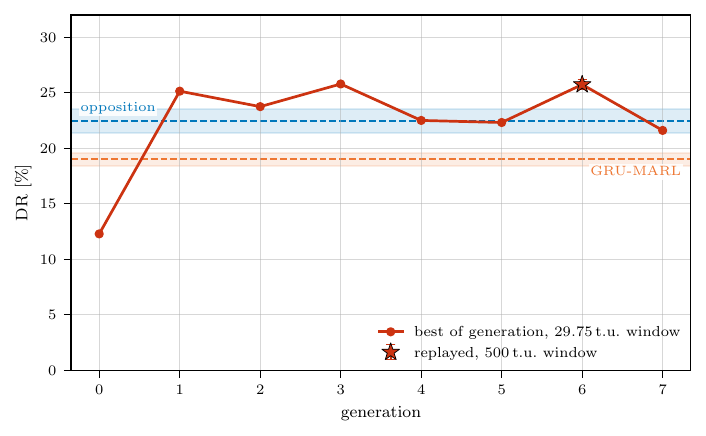}
  \caption{Refinement on the evaluation domain. Circles give the best fitness of
  each generation, scored over the $29.75$ t.u.\ rollout each candidate is given
  during the search and shifted by a single constant so that the deployed
  candidate reads the drag reduction measured for it over $500$ t.u. Stars give
  that measured drag reduction for the candidates replayed over the long window,
  which is the window every number quoted in the text uses. Dashed lines mark
  opposition control and the transferred GRU-MARL policy on the same domain, with
  bands of one block-averaged standard error. Most of the gain is present after
  one generation.}
  \label{fig:escurve}
\end{figure}

Refinement on the evaluation domain recovers it and more. Table~\ref{tab:dr}
collects the drag reduction, the friction velocity and the effective friction
Reynolds number of the four flows. The refined controller reaches $25.7\%$,
above the $22.5\%$ of opposition control and $6.7$ points above its own unrefined
parent.
The recovery is not gradual. Figure~\ref{fig:escurve} shows the fitness and the
population-mean drag reduction against generation: the population mean rises from
$3.2\%$ at the first generation to $21.7\%$ by the sixth and then flattens, while
the mean penalty falls monotonically from $0.116$ to $0.076$, so the later
generations are buying smoothness rather than drag. Most of the gain is present
after a single generation, which is consistent with the reading that the
minimal-box stage had already located the right region of parameter space and the
refinement is correcting the policy to the flow it now sees rather than searching
afresh.

\begin{table}[tb]
\caption{Drag reduction, friction velocity and effective friction Reynolds number
for the four flows on the evaluation domain. Drag reduction is measured against
the non-actuated pressure gradient, over $500$ convective time units for the two
learnt controllers, thirty-two eddy turnovers, and $100$ for the remaining
cases.}
\label{tab:dr}
\setlength{\tabcolsep}{14pt}
\begin{tabular}{lccc}
\toprule
Case & $\DR$ & $\ufric$ & $\Retau^{\mathrm{eff}}$ \\
\midrule
non-actuated & $0$       & $0.0640$ & $183.8$ \\
opposition   & $22.5\%$ & $0.0564$ & $161.7$ \\
GRU-MARL     & $19.0\%$ & $0.0576$ & $165.3$ \\
ES           & $25.7\%$ & $0.0552$ & $158.3$ \\
\bottomrule
\end{tabular}
\end{table}

The short scoring window of \S\ref{sec:es} deserves a check, since $29.75$ time
units is under two eddy turnovers, and a ranking made over such a window can
reflect the realisation rather than the policy. Re-evaluating the selected
candidate over $500$ time units, thirty-two eddy turnovers, returns $25.7\%$
against the $25.8\%$ its scoring window reported.
The ranking held in this case. It need not in general, and the margin between
candidates within a generation is comparable to the spread we measure across
windows, which is a limitation of the procedure rather than of the result.

Having reached comparable drag reductions by different routes, the remainder of
the paper asks whether the underlying flow states coincide, and in particular what
the refinement stage changed.

\subsection{Where the friction is removed}\label{sec:res-fik}

\begin{figure}[tb]\centering
  \figmaybe{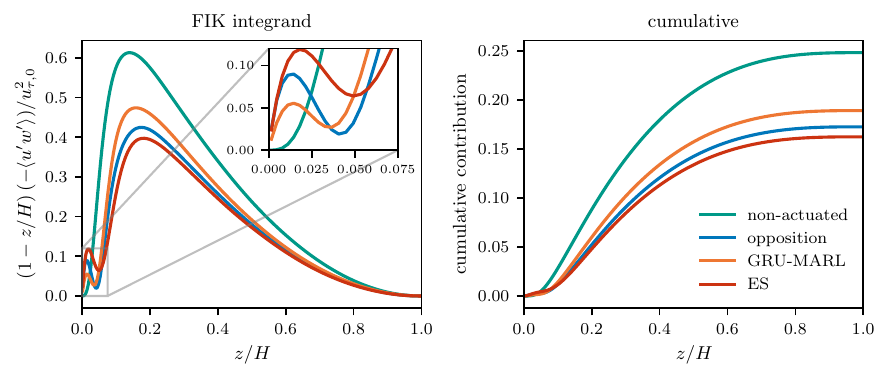}
  \caption{Wall-normal decomposition of the skin friction: the weighted,
  normalised Reynolds stress (left) and its cumulative integral (right) against
  $z/H$. The area between the non-actuated curve and a controlled one is that
  controller's drag reduction.}
  \label{fig:fik}
\end{figure}

Figure~\ref{fig:fik} decomposes the skin friction across the channel with the
identity of Eq.~\eqref{eq:fik}, plotted against $z/H$ and scaled for every case by
the single friction velocity of the non-actuated flow, so that the area between
the non-actuated curve and a controlled one is that controller's drag reduction.
All three controlled flows carry a smaller weighted Reynolds stress than the
non-actuated flow across the buffer layer and the region above it, and the
cumulative curves show the reduction accruing over the same range of $z/H$. The
relative split between layers is similar across the controllers, the buffer layer
carrying about a quarter of the weighted stress in each case, while the turbulent
fraction of the mean-energy budget falls from $0.41$ in the non-actuated flow to
between $0.34$ and $0.35$ under control.

Table~\ref{tab:decomp} gives the split.

\begin{table}[tb]
\caption{Fraction of the weighted Reynolds-stress integral of
Eq.~\eqref{eq:fik} carried by the viscous sublayer, the buffer layer and the
region above it, and the turbulent fraction of the mean-energy budget in the
decomposition of Renard and Deck.}
\label{tab:decomp}
\setlength{\tabcolsep}{9pt}
\begin{tabular}{lcccc}
\toprule
case & viscous & buffer & log$+$ & turb.\ frac.\ (RD) \\
\midrule
non-actuated & 0.002 & 0.259 & 0.720 & 0.411 \\
opposition & 0.011 & 0.249 & 0.720 & 0.348 \\
GRU-MARL & 0.006 & 0.250 & 0.725 & 0.339 \\
ES & 0.017 & 0.254 & 0.708 & 0.353 \\
\bottomrule
\end{tabular}
\end{table}

The decomposition therefore locates the reduction but does not on its own
distinguish the controllers, and in particular it does not distinguish the policy
before refinement from the policy after it. The difference lies in how the
near-wall stress is organised, which the remaining diagnostics address.

\subsection{Reynolds-stress reorganisation and anisotropy}\label{sec:res-stress}

\begin{figure}[tb]\centering
  \figmaybe[0.49\linewidth]{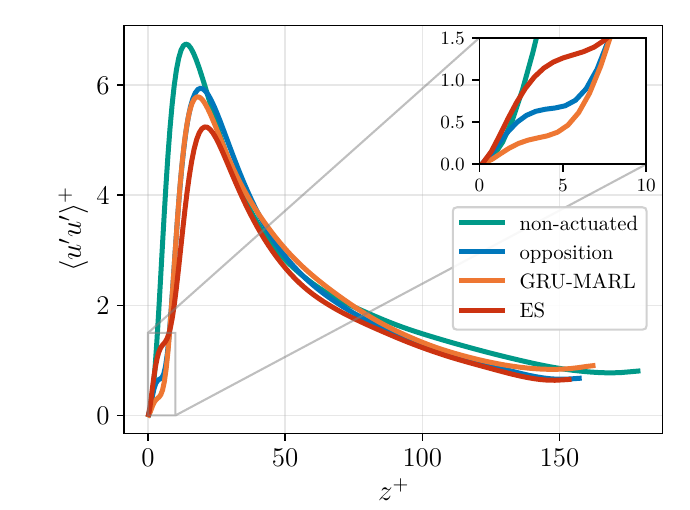}\hfill
  \figmaybe[0.49\linewidth]{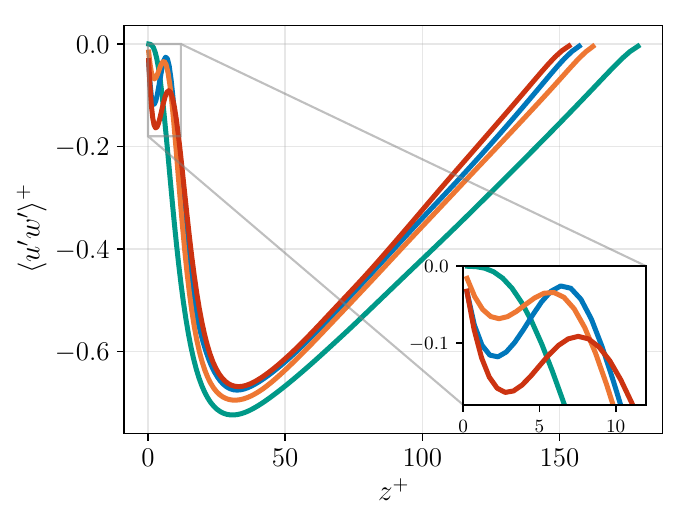}
  \caption{Streamwise normal stress $\avg{u'u'}^{+}$ (left) and Reynolds shear
  stress $\avg{u'w'}^{+}$ (right), each renormalised by the controller's own
  post-control friction velocity.}
  \label{fig:stresses}
\end{figure}

Figure~\ref{fig:stresses} shows the streamwise normal stress $\avg{u'u'}^{+}$ and
the Reynolds shear stress $\avg{u'w'}^{+}$, each normalised by its own friction
velocity. The shear stress, which carries the drag, follows opposition for both
learnt controllers, yet their streamwise normal stress differs from that of every
other case, opposition included. Reaching an opposition-like shear stress through
a streamwise fluctuation field that no benchmark reproduces is a property the two
learnt controllers share, and it is inherited from the architecture and the
gradient-based stage rather than produced by the refinement.

\begin{figure}[tb]\centering
  \figmaybe{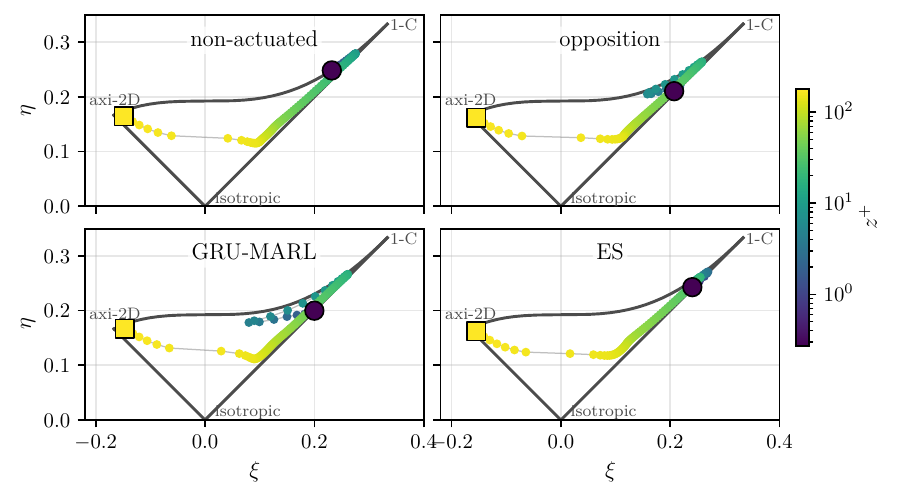}
  \caption{Anisotropy-invariant trajectories through the wall-normal coordinate
  for the four flows.}
  \label{fig:aniso}
\end{figure}

Figure~\ref{fig:aniso} maps the same information onto the plane of the
anisotropy invariants $\xi$ and $\eta$, and the controlled flows do not leave the
wall from the same place. Opposition and GRU-MARL start from very nearly the same
point, whereas the refined controller starts closer to the one-component limit, so
its departure towards one-dimensional structure is already the stronger one where
the actuation is applied. Above the wall the paths separate. GRU-MARL turns
towards the axisymmetric two-component state and sits near it through the buffer
layer. Opposition sets out in the same direction but holds a higher $\xi$ and does
not arrive. The refined controller does not raise $\xi$ at all and stays close to
the state it began in, so the configuration it establishes at the wall is carried
up through the buffer layer instead of relaxing towards the shape the other two
adopt. Opposition displaces the state along one invariant; the refined controller
displaces it in both, which is the signature of a wall actuation coupled to the
buffer layer through more than a single component.

\subsection{What refinement changed in the control law}\label{sec:res-policy}

\begin{figure}[tb]\centering
  \fignat{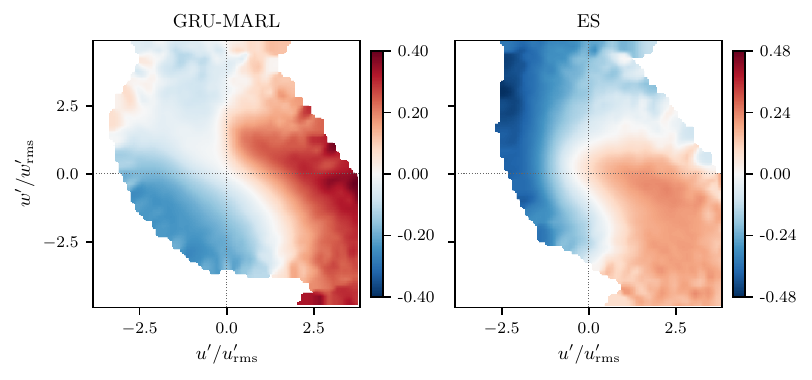}
  \caption{Conditional policy maps for the two learnt controllers: mean actuation
  over $(u',w')$ bins at the detection plane, before and after refinement.}
  \label{fig:policymap}
\end{figure}

The diagnostics so far describe the flow. Because the two learnt controllers
differ only by the refinement stage, it is also possible to ask what refinement
did to the control law itself. Figure~\ref{fig:policymap} conditions the mean
actuation on the local streamwise and wall-normal fluctuations at the detection
plane, for the policy before and after refinement. Opposition, by construction,
produces a map that varies with $w'$ alone. The unrefined policy departs from that
only weakly. After refinement the map varies principally with $u'$ and much less
strongly with $w'$: across the learnt controllers the correlation of the actuation
with $u'$ is about $0.5$ to $0.6$, two to four times stronger in magnitude than
its correlation with $w'$, and the separation is widest for the refined policy.

The refinement therefore did not simply scale up an existing strategy. It moved
the policy from a weak wall-normal coupling towards a strong streamwise one, that
is, from something resembling opposition applied imperfectly towards a controller
that tracks the streaks. That streamwise coupling is not itself a new control
principle, since feedback laws aligned with the streamwise wall shear have been
derived analytically from the suppression of the near-wall Reynolds shear
stress\cite{fukagata2004suboptimal} and proposed from resolvent
analysis;\cite{kawagoe2019resolvent} what is new here is that it emerged from
refinement on the target domain, in a policy that did not exhibit it before, with
no change of architecture, objective family or sensing station. We describe the
spatial relationship only, and do not address the order in which the actuation and
the structures influence one another.

\subsection{Near-wall structure and scale selectivity}\label{sec:res-structure}

\begin{figure}[tb]\centering
  \figmaybe{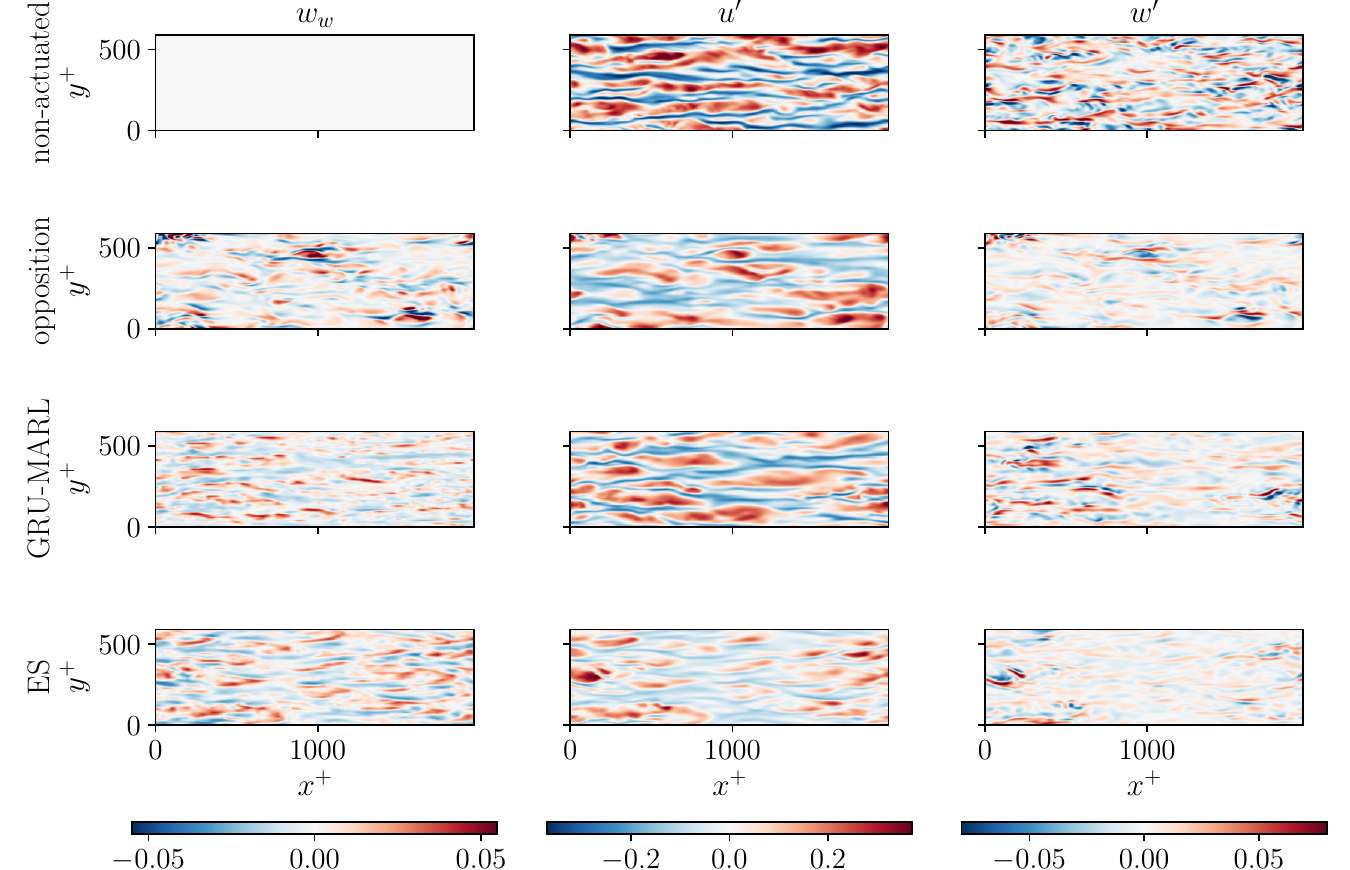}
  \caption{Instantaneous fields for the four flows: the wall actuation $\ww$ at
  the wall, and the streamwise and wall-normal fluctuations $u'$ and $w'$ at the
  detection plane $\zd^{+}\simeq14$.}
  \label{fig:structure}
\end{figure}

Figure~\ref{fig:structure} shows instantaneous fields of the near-wall velocity
and of the wall actuation for each controller. Opposition sets the wall velocity
to the negative of the sensed wall-normal fluctuation, so its actuation follows
the footprint of the sweeps and ejections at the detection plane by construction.
The refined actuation instead aligns with the streamwise low- and high-speed
streaks, as the conditional maps of \S\ref{sec:res-policy} anticipate.

\begin{figure}[tb]\centering
  \figmaybe{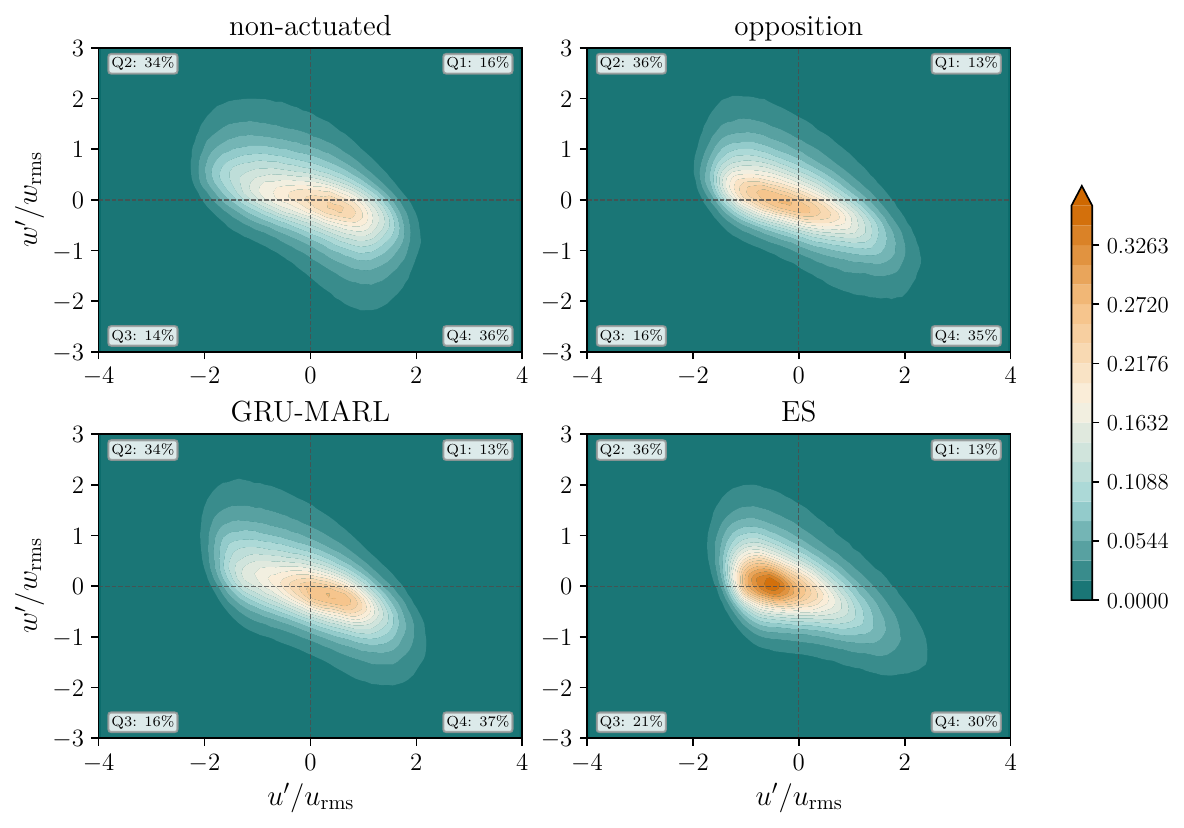}
  \caption{Joint probability density of $(u',w')$ at the detection plane for the
  four flows, including the non-actuated reference. Boxes give quadrant occupation
  fractions.}
  \label{fig:quadrant}
\end{figure}

The joint probability density of $(u',w')$ at the detection plane, in
Fig.~\ref{fig:quadrant}, shows how the controllers redistribute the
momentum-carrying events. In the non-actuated flow the ejections and sweeps
contribute almost equally. Under every controller the sweep contribution exceeds
the ejection contribution, so the ejections are weakened preferentially, and the
effect is strongest for the refined controller, whose ejection contribution to the
inner-scaled Reynolds shear stress falls to about $-0.10$ from $-0.32$ in the
non-actuated flow.
These are contributions to the shear stress and not the quadrant occupation
probabilities annotated in the figure. Table~\ref{tab:quadrant} collects them.

\begin{table}[tb]
\caption{Quadrant contributions to the inner-scaled Reynolds shear stress at the
detection plane $\zd^{+}\simeq14$, each case scaled by its own friction velocity.
Q2 are ejections and Q4 sweeps.}
\label{tab:quadrant}
\setlength{\tabcolsep}{9pt}
\begin{tabular}{lccccc}
\toprule
case & Q1 & Q2 (ej.) & Q3 & Q4 (sw.) & $\langle u'w'\rangle^+$ \\
\midrule
non-actuated & +0.055 & -0.323 & +0.042 & -0.336 & -0.562 \\
opposition & +0.028 & -0.153 & +0.020 & -0.218 & -0.322 \\
GRU-MARL & +0.027 & -0.202 & +0.027 & -0.222 & -0.371 \\
ES & +0.026 & -0.101 & +0.021 & -0.175 & -0.229 \\
\bottomrule
\end{tabular}
\end{table}

\begin{figure}[tb]\centering
  \fignat{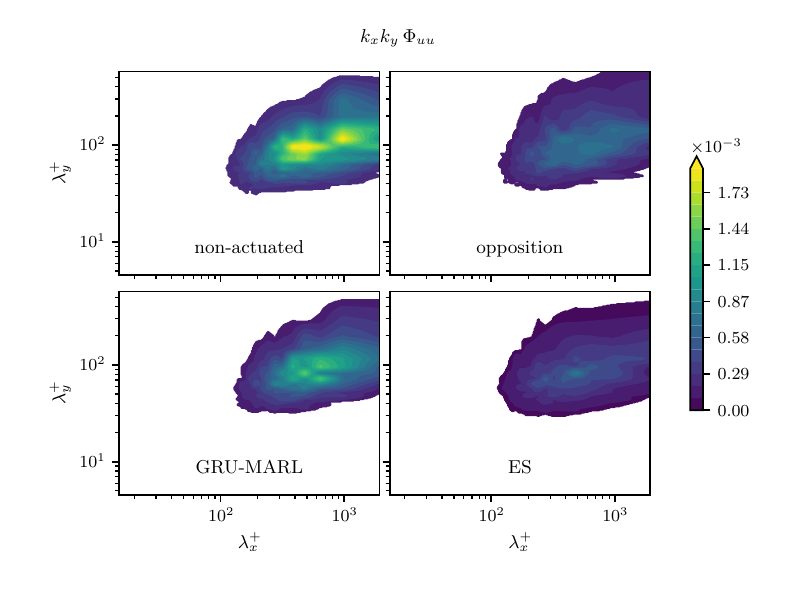}
  \caption{Premultiplied spectra of the streamwise velocity fluctuation near the
  wall, at $\zd^{+}\simeq14$, for the four flows. The streak peak near
  $\lambda_x^{+}\simeq700$ survives under GRU-MARL and is redistributed towards
  shorter streamwise wavelengths after refinement.}
  \label{fig:spectra_uu}
\end{figure}

\begin{figure}[tb]\centering
  \fignat{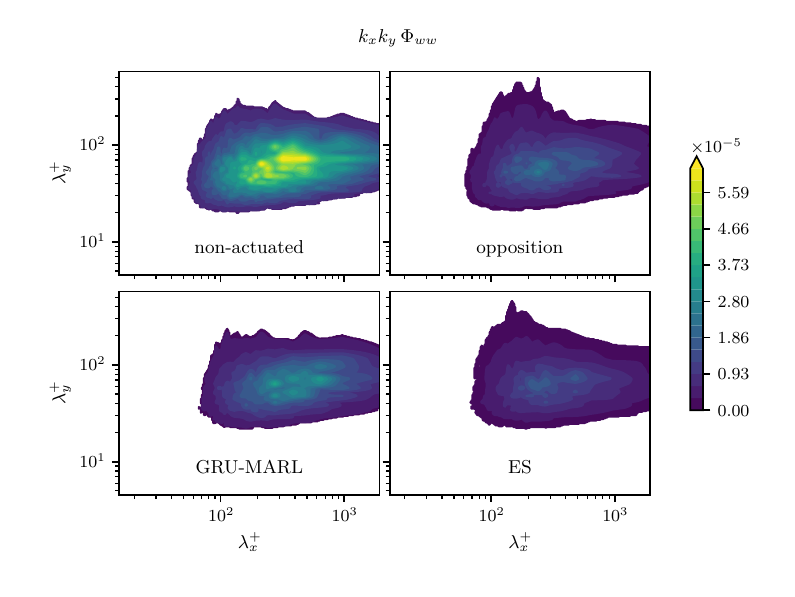}
  \caption{Premultiplied spectra of the wall-normal velocity fluctuation near the
  wall, at $\zd^{+}\simeq14$, for the four flows. The controlled flows are modified
  around the spanwise streak wavelength $\lambda_y^{+}\simeq100$.}
  \label{fig:spectra_ww}
\end{figure}

Figures~\ref{fig:spectra_uu} and~\ref{fig:spectra_ww} show the premultiplied
spectra of the streamwise and wall-normal velocity fluctuations near the wall. The energy of the controlled
flows is modified at the spanwise wavelengths of the near-wall streaks, around
$\lambda_y^{+}\simeq100$,\cite{delalamo2004scaling} and the learnt controllers
concentrate their effect there.
The streamwise streak energy, peaked near $\lambda_x^{+}\simeq700$ in the
non-actuated and GRU-MARL flows, is shifted to shorter wavelengths under
opposition and most strongly redistributed towards small scales by the refined
controller, whose long streaks are broken into shorter fragments.

\subsection{Smoothness and statistical steadiness}\label{sec:res-smooth}

\begin{figure}[tb]\centering
  \fignat{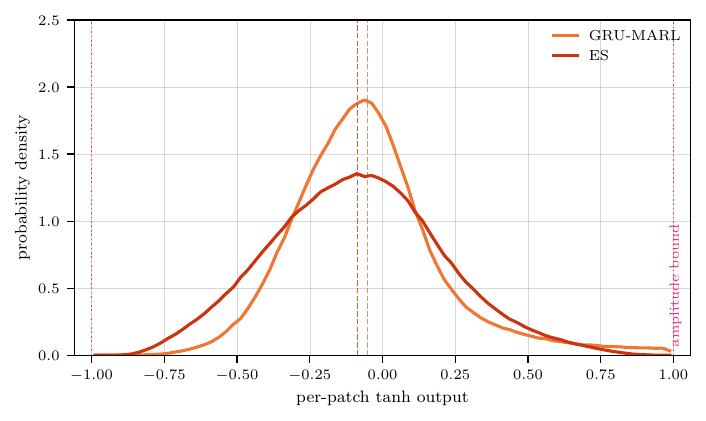}
  \caption{Distribution of the raw per-patch decisions, the network output before
  smoothing, projection and scaling, for the two learnt controllers. Dashed lines
  mark each distribution's mean and dotted lines the amplitude bound at $\pm1$.
  Neither controller accumulates weight against the bound.}
  \label{fig:smooth-actions}
\end{figure}

The comparison above is meaningful only if the refined actuation controls the
turbulence rather than settling into a fixed pattern set by the box, which is the
failure mode that motivated the penalties in Eq.~\eqref{eq:fitness}.
Figure~\ref{fig:smooth-actions} shows the distribution of wall-actuation values
for each controller. It is smooth and stays away from the amplitude bound, with a
root-mean-square of $0.014$ and a negligible saturated fraction.
The streamwise-averaged actuation remains statistically stationary and does not
lock into a standing wave. The actuation is therefore an unsaturated response that
tracks the evolving flow. Appendix~\ref{app:unreg} shows what an objective scored
on the pumping power alone favours instead.

\section{Discussion}\label{sec:discussion}

A gradient-free refinement stage run on the target domain recovers the drag
reduction that a minimal-box multi-agent policy loses when it is transferred, and
carries it past opposition control. The mechanism of the gain is not subtle. The
search scores each complete episode by a single number and never assigns a
separate reward to each wall patch, so the zero-net-mass constraint that links the
patches becomes part of how the actuation is formed rather than a source of
interference in the optimisation, and the controller is tuned on the same flow it
will drive. The cost is a population of rollouts per generation, which are
independent and parallelise cleanly, and which here amounted to $19$ hours on one
GPU for eight generations.

The two stages are complementary rather than competing, and the result should not
be read as an ablation in which the evolution strategy beats gradient descent.
Gradient training on the large domain degrades as the patch count grows, for the
reasons set out in \S\ref{sec:intro}, and each episode costs a direct numerical
simulation. A run of that kind on a domain of this size takes some three hundred
to a thousand episodes, collected one after another because each is scored against
a policy the preceding episode has already altered, and in practice it also takes
hyperparameters adjusted while the run is under way, since the training otherwise
settles either at a drag reduction well below what the architecture can reach or
on a saturated policy. The $192$ episodes of the refinement are mutually
independent, and they were completed in $19$ hours on a single GPU.
A strategy started from random parameters on the large domain would
spend its budget locating the region that the minimal-box training already found.
What the strategy is good at is moving around a good point in parameter space,
scoring whole episodes, and doing so across rollouts that parallelise. The
minimal-box stage supplies the point; the refinement supplies the correction to
the flow that matters. The scope of the claim follows from that design: the final
controller inherits the region of parameter space it started from, and what has
been established is that this region can be improved on the domain of interest.
Whether a distant region holds a better controller is untested, and testing it
would require the from-scratch search the two-stage procedure exists to avoid.

The comparison of the flow before and after refinement gives the physical content
of the gain. Opposition and the refined controller remove comparable amounts of
skin friction, yet the diagnostics place their flows differently through the
buffer layer, with different Reynolds-stress profiles when each is scaled by its
own friction velocity and different anisotropy. The clearest separation is in the
streaks. The instantaneous fields and the streamwise spectra (Fig.~\ref{fig:spectra_uu})
show opposition
mainly lowering the streak amplitude, whereas the refined controller breaks the
long streaks into shorter fragments and moves the streamwise energy to smaller
scales. Neither the effect nor its value is new: a suboptimal control built on the
suppression of the near-wall Reynolds shear stress breaks the long streaks in much
the same way,\cite{fukagata2004suboptimal} at a drag reduction its authors
describe as modest. That the same reorganisation appears here at a larger saving
suggests the mechanism is not exhausted, and that a control law which engages it
more completely, or engages it alongside a second one, remains a route to a more
sustained reduction. Spanwise forcing produces a comparable lateral
reorganisation of the streaks,\cite{touber2012spanwise} though through an open-loop
rather than a sensed actuation. That two controllers reaching similar drag reduction act so differently
on the coherence of the near-wall cycle indicates different mechanisms,
distinguished by the flow topology rather than by the drag figure alone.

The change in the control law is the part attributable to the refinement itself,
because the architecture, the objective family and the sensing station are
unchanged across it. Refinement moved the actuation from a weak coupling to the
wall-normal velocity, which is what opposition acts
on,\cite{choi1994opposition,hammond1998oppositionmech} towards a strong coupling
to the streamwise fluctuation. Control laws aligned with the streamwise wall shear
have been derived before from the suppression of the near-wall Reynolds shear
stress\cite{fukagata2004suboptimal} and from resolvent
analysis,\cite{kawagoe2019resolvent} so the strategy the refined policy converged
on is not without precedent; what the present result adds is that a learnt policy
moves towards it of its own accord once it is optimised on the domain it will run
on, and does not while it is optimised on a smaller one.

Two limitations bound the result. The sensing plane sits inside the flow at
$\zd^{+}\simeq14$, which is not measurable in a practical installation, and
estimating it from wall quantities in the presence of control is itself an open
problem that a companion study addresses. Holding it fixed here is deliberate,
since it is what makes the before-and-after comparison controlled. Second,
everything reported is at a single Reynolds number. Performance built on the near-wall cycle is known to
deteriorate as $\Retau$ rises, because the motions carrying the friction move away
from the wall,\cite{gatti2016stwhighre,marusic2021energypath,zhou2025drlhighre}
and nothing here tests whether the refined controller is an exception. What can be
said is that the refinement stage, unlike the minimal box, imposes no obstacle of
principle to being repeated at another Reynolds number or in another geometry,
because it needs only a solver, an objective and enough parallel rollouts.

\section{Conclusions}\label{sec:conclusions}

A recurrent multi-agent wall controller trained on a minimal flow unit at
$\Retau\simeq180$ delivers between $24\%$ and $27\%$ drag reduction on the box it
was trained in and $19.0\%$ on a channel sixteen times larger in wall-parallel
area, at matched resolution, Reynolds number, patch size and actuation cadence.
Refining its $48\,833$ parameters with an evolution strategy on the larger channel,
scoring whole episodes against a regularised drag-reduction objective, raises that
to $25.7\%$, above the $22.5\%$ of opposition control, with most of the gain
present after a single generation.

Because the policies before and after refinement share an architecture, a training
history and a sensing station, the difference between the two controlled flows is
attributable to the refinement. It changes where the flow sits in the
anisotropy-invariant triangle, how the momentum-carrying events are redistributed
at the detection plane, and how the streamwise energy is spread across scales, and
it changes the control law from a weak coupling to the wall-normal velocity to a
strong coupling to the streamwise fluctuation.

The wider point concerns how learnt wall controllers should be assessed. A
controller optimised in a box smaller than the flow it will drive may be
optimising the box, a hazard documented for evolutionary optimisation of compliant
walls two decades ago\cite{fukagata2008compliant} and present in a different form
whenever a reinforcement-learning policy is trained in a minimal flow unit. The
remedy used here is to move the last stage of the optimisation onto the domain the
controller will be judged on, which a gradient-free method makes possible without
retraining from scratch.

\begin{acknowledgments}
G.M.C.\ acknowledges support from the EPSRC Doctoral Training Partnership, grant
EP/W524608/1. M.P.C.\ was supported by the Horizon Europe Marie
Sk\l{}odowska-Curie Doctoral Network SCALE, grant agreement No.\ 101120014. The
simulations and the evolution-strategy search were run on an NVIDIA GB10 Spark.
\end{acknowledgments}

\section*{Author declarations}

\subsection*{Conflict of interest}

The authors have no conflicts to disclose.

\subsection*{Author contributions}

\textbf{Giorgio M. Cavallazzi}: Writing -- review \& editing (equal); Writing --
original draft (lead); Visualization (lead); Validation (lead); Software (lead);
Resources (lead); Methodology (equal); Investigation (lead); Data curation (lead);
Conceptualization (equal). \textbf{Miguel P\'erez Cuadrado}: Writing -- review \&
editing (equal); Visualization (supporting); Methodology (equal); Formal analysis
(supporting); Conceptualization (equal). \textbf{Alfredo Pinelli}: Writing --
review \& editing (equal); Supervision (lead); Project administration (lead);
Methodology (equal); Conceptualization (equal).

\section*{Data availability}

The data that support the findings of this study are openly available in GitHub at
\url{https://github.com/gmcavallazzi/CaNS_GRU-MARL}. The repository holds the
modified CaNS flow solver, the multi-agent reinforcement-learning training
framework, the evolution-strategy driver and the trained policies, so every
controller reported here can be re-run and evaluated.

\appendix

\section{What the unregularised objective converges to}\label{app:unreg}

Section~\ref{sec:es} states that optimising the net energy saving on its own,
without the four penalties of Eq.~\eqref{eq:fitness}, does not produce a usable
controller. This appendix records that run.

The search is identical to the one reported in the main text in every respect
except the objective. It starts from the same warm start, uses the same population
of $24$ mirrored candidates, the same perturbation scale $\sigma=0.02$, the same
Adam settings, the same $29.75$ time-unit scoring window and the same evaluation
domain, and it runs for the same eight generations. The only change is that the
fitness is the bare net saving $(\Eps_0-\Eps)/\Eps_0$ rather than drag reduction
less the four penalty terms.

Within the scoring window the search succeeds, and by a wider margin than the
regularised one. The best net saving climbs from $0.177$ in the first generation
to $0.286$ by the eighth, against $0.243$ for the regularised search, and no
candidate at any generation trips the relaminarisation guard. Over the window it
was given, the unpenalised search scores higher.

\begin{table}[tb]
\caption{The selected candidate of each search, evaluated over the same $100$
time-unit window on the same domain. The unregularised policy reports the higher
drag reduction and the lower net saving, because it delivers three times the wall
power; its streamwise momentum balance also closes three times less well.}
\label{tab:unreg}
\setlength{\tabcolsep}{12pt}
\begin{tabular}{lcc}
\toprule
 & regularised & unregularised \\
\midrule
score in the $29.75$ t.u.\ window & $0.243$             & $0.286$             \\
$\DR$                             & $26.1\%$            & $26.3\%$            \\
$(\Eps_0-\Eps)/\Eps_0$            & $25.9\%$            & $25.8\%$            \\
$\Ww$                             & $6.9\times10^{-6}$  & $2.1\times10^{-5}$  \\
$\Epump$                          & $3.03\times10^{-3}$ & $3.02\times10^{-3}$ \\
momentum residual                 & $1.5\%$             & $3.9\%$             \\
\bottomrule
\end{tabular}
\end{table}

The advantage disappears over a longer window. Table~\ref{tab:unreg} sets the two
selected candidates side by side over a full $100$ time units. The unregularised
policy returns a drag reduction of $26.3\%$ against $26.1\%$ and a net saving of
$25.8\%$ against $25.9\%$. Its net saving is the quantity it was optimising, and
it falls from $0.286$ over the scoring window to $0.258$ over the long one. The
regularised candidate rises over the same comparison, from $0.243$ to $0.259$.

The wall power accounts for the difference. The unregularised policy delivers
$\Ww=2.1\times10^{-5}$ against $6.9\times10^{-6}$, three times as much, so it buys
its pressure-gradient reduction by pushing more energy through the wall at the
same pumping power. Its streamwise momentum balance closes three times less well,
$3.9\%$ against $1.5\%$, consistent with an actuation carrying large-scale
structure that the mean balance must absorb. The actuation itself organises into a
large-scale travelling pattern, of the kind reported for wall controllers scored
on quantities insensitive to how the actuation is
distributed.\cite{cavallazzi2026rewardhacking}

Equation~\eqref{eq:fitness} takes its form for this reason. An objective charging
only the net energy saving cannot separate a controller from a forcing over a
window short enough to be affordable, because the covariance $\avg{\ww\,p}$ that
would expose the difference is a small number set against a pumping power three
orders of magnitude larger, and its estimate over thirty time units is
correspondingly noisy. The temporal, spatial, zero-mean and amplitude penalties
leave the measured quantity untouched. They hold the search in the region where
the actuation is smooth, unsaturated and genuinely closed-loop, where a score
obtained over a short window carries over to a long one.

\end{document}